# Examining the impacts of privacy awareness on user's self-disclosure on social media.

Kijung Lee[1] & Prudence Attablayo[1]

[1] University of Cincinnati, Cincinnati OH 45221, USA
`kijung.lee@uc.edu`

**Abstract.** This research aims to investigate the impact of users' privacy awareness on their self-disclosing behavior. Our primary research question is to investigate how young social media users feel about the benefits and risks of disclosing themselves on social media and how risk-benefit awareness influences the assessment of their self-disclosure. Based on the data we recorded, the factor analysis, and three-way ANOVA, we conclude that users who know more about privacy benefits share more on social media (F= 36.291; df 1; sig < .001) while those who know less about the benefits disclose less on social media. According to the analysis, users who know more about self-disclosure risks share less on social media (F= 7.001; df 1; sig < .001). Users disclose less information on social media platforms based on the different levels of their risk perceptions (df 3, F=.715, sig < 0.5). This indicates that risks on social media platforms vary to some degree. We saw that people's sharing habits based on their levels of risk, benefits, and social media platforms can vary. One thing that remained certain was users' main benefit for engaging and disclosing on social media is their need to stay in touch with friends and their need for community. On the flip side, the main risk was the need not to be impersonated and misunderstood by people. Based on a simple frequency analysis of the open-ended questions we asked In our data collection, the most highlighted words in our responses were "people" and "friends". These were the two main words that stood out in all the data we collected concerning the benefits, risks, and intention to self-disclose.

## 1 Introduction

As the distinctions between our online and physical lives continue to become hazier, privacy concerns on social media are more crucial than ever. With a single click, you can instantly broadcast your opinions, activities, and even private information to the world. It's vital to think about the effects of your online conduct as you navigate the digital world and to take precautions to keep yourself and your personal information safe from public scrutiny.

With the help of social networking services, there has been a significant increase in user engagement in recent years. TikTok alone is one of the most used platforms with over 750 million users. People in today's world are becoming more open to sharing personal information and with the help of our current digital culture, this has been made possible using the convenience of internet services Cavusoglu, Phan, Cavusoglu, and Airoldi, (2016). However, as connectivity increases with smart mobile devices and social media use, so does the risk of data breaches Crossler and Bélanger, (2019), Gerhart and Koohikamali, 2019, Gu et al., (2017). In all this, there is a thin line between user-sharing behaviors and privacy awareness. There have been many attempts to define privacy in various research papers related to privacy in e-commerce, social networking websites, and other fields - In its broadest sense, privacy is often associated with the "right to be alone" Warren & Brandeis, (1890). Klopfer & Rubenstein, (1977) views privacy as a regulatory process designed to selectively control access to external stimuli oneself or the flow of information to others. It can also be viewed as a claim that can be redeemed for more value. For so long the concept of privacy has been viewed from the lens of privacy calculus with the assumption that: users make rational decisions based on their preconceived perception of the benefits and risks they will gain from sharing their private information. However, users have not been questioned on their actual knowledge of these benefits and risks. This research examines privacy awareness based on users' awareness of benefits and risks and how it translates into their self-disclosure intentions.

Over the past two years, we have seen a rise in social media user engagement. The lockdown led to many people sharing a portion of their private lives and time on social media; out of these practices, some good and bad things have emerged. In terms of good things we have seen roles like social media engagement manager, influencer, and content strategist amongst others become full-time jobs for others. One other good thing is the new way of making friends on social media. Through the lockdown, people found others on social media with similar likes and out of several conversations have grown friendships. In terms of the bad things, we have seen an increase in depression, cyberbullying, and crime influenced by inadequate social media privacy sharing awareness. The main objective of this research is to investigate how younger users feel about the benefits and risks of disclosing themselves on social media and how the younger users' self-disclosure is influenced by the risk-benefit assessment of self-disclosure based on various social media platforms. In the subsequent sections of this paper, we discuss the theoretical background to privacy calculus and privacy awareness in the literature review. From there, we discuss the three main categories of benefit and risk awareness we identified. The methodology section comes next where we discuss our data collection, population, and data analysis and then we conclude with our discussion and conclusion.

3## 2 Literature Review

### 2.1 Privacy calculus

Privacy calculus is a way of understanding the privacy and security trade-offs of a given technology, system, or organization. The term Privacy Calculus, originally known as the "calculus of behavior" assumes that people will divulge personal information when the perceived benefits outweigh the possible costs Laufer & Wolfe, (1977). Knijnenburg et al. (2017) say that the decision to share data is characterized by privacy calculus as a balance between the benefits and risks of disclosing personal information. Nevertheless, numerous researchers have discovered that these decisions are often not calculated at all and that users devise strategies to work around disclosing their information (Acquisti, 2005). Critics claim that the privacy calculus is overly simplistic and does not account for the realities of individuals' decision-making processes when making a decision on whether to disclose personal information Dinev, T., & Hart, P. (2006). Over the years this concept has been studied through different approaches like game-simulated scenarios to see how users act when they have to make a privacy-related decision. Other researchers have also used the field of machine learning. With these studies came the concept of user-tailored privacy. This concept deals with some of the moral issues that other privacy solutions pose. Privacy calculus involves weighing the costs and benefits of sharing the information and the level of trust the individual has in the recipient of the information. Other critics also claim that the privacy calculus makes several unverified presumptions, such as the notion that individuals have constant and stable privacy preferences, or that people can properly estimate the risks and benefits of their privacy decisions.

The primary criticism about the privacy calculus is that most regression models designed and tested in the existing privacy calculus are based on the assumption that users can make rational decisions about self-disclosure when they have no experiential information about the impact of self-disclosure. So, in the context of this thesis, we want to include the awareness (knowledge) of the users about the benefits and risks and how self-disclosure is managed based on the knowledge.

There are several factors that can influence an individual's privacy calculus, including the perceived value of the information being shared, the potential risks and consequences of disclosure, and the individual's level of control over the information. Cultural and societal factors can also influence privacy calculus, such as the legal and social norms surrounding privacy in a particular location.

The depth and diversity of life that comes with progressing society has forced some withdrawal from the world, which has caused man to become more sensitive to publicity as a result of culture's refining influence and has made solitude and seclusion more important to the individual Warren & Brandeis, (1890). Knijnenburg et al.,( 2017) say that the decision to share data is characterized by privacy calculus as a balance between the benefits and risks of disclosing personal information. People have various reasons for sharing information online some of which are for their personal enjoyment and satisfaction Lee, Goh, Chua, & Ang, (2010). This paper has catego-rized and will explore the identified three major benefits and risks of privacy awareness.

The privacy calculus is used prescriptively in user-tailored privacy, and machine learning algorithms that use the risk/benefit trade-off as an objective function. Existing research has shown that most users make heuristic-based privacy decisions as compared to making rational decisions. Previously it can be argued that the privacy calculus approach has viewed privacy from users' retrospective and rational perspectives. In this research, we are using the concept of user awareness to more precisely measure users' awareness of benefits and risks and how it translates into their self-disclosure intentions. Some gaps that have been identified in previous research are

Limited knowledge of how the layout and user experience of online platforms affect people's privacy calculations Cranor & Doolin, (2011). Lack of study on children's and teens' privacy calculus Lenhart et al., (2010). A lack of comprehension of the function that emotions play in the privacy calculus (Acquisti & Grossklags, 2005)

## 2.2   Privacy awareness

The more aware individuals are of the possible hazards and dangers involved with revealing personal information online, the less likely they are to engage in such behavior Joinson, A. N. (2008). Considering one's own privacy rights and how to safeguard them, as well as the privacy policies and practices of organizations like social media firms, are all part of privacy awareness. On social media, this entails being knowledgeable about how the platform gathers, utilizes, and disseminates personal data and taking precautions to safeguard one's personal data, such as modifying privacy settings and exercising caution when sharing information publicly. It also entails being informed of the dangers and repercussions of disclosing personal information online. For this research, we identified our target audience's three main most used social media applications. These applications are Snapchat, Tiktok, and Instagram. We saw the disparity in self-disclosure level based on the platform's features. It has previously been identified that some platforms enable more user sharing than others. For platforms like Twitter which is mostly text-oriented we and In this section, we delve into the three most common benefits and risks we identified during this research.

**Benefit awareness**
There are various reasons that contribute to why people share information online, some of which are for their personal enjoyment and satisfaction Lee, Goh, Chua, & Ang, (2010), social benefits which include but are not limited to social capital, relationship management, personal branding and for societal benefits - where societal benefits are with regards to adding value to the society using social media as a platform.

We investigated previous research and came to the conclusion that these are the primary benefits of social media. These benefits are Personal Benefits, Social Benefits, and Societal Benefits.

**Personal benefits**
Williamson, P., Stohlman, T., Polinsky, H. (2017) A Survey of Self-disclosure Motivations on Social Media," looks at the reasons people disclose themselves on social media. This study presents the results of a survey of social media users who were questioned about the reasons they gave for disclosing personal information online. According to research, sustaining relationships, getting feedback and validation from others, and projecting a specific image to others are the three most frequent reasons for self-disclosure on social media. The urge to express oneself, they want to connect with friends and family, and the desire to find others who have similar interests were among the other factors mentioned in the survey. Understanding the motivations behind self-disclosure on social media is crucial to comprehend social media's function in people's lives. Self-disclosure on social media is driven by a complex interplay of societal and personal elements. Consequently, it is critical.

**Societal benefits**
Taneja, A., Vitrano, J., & Gengo, N. J. (2014) talk about how self-disclosure on social networking sites has changed from being a personal estimate of privacy to considering social ties and relationships. It advises people to consider the advantages and disadvantages of disclosing personal information online in light of their social networks and the social norms of their online communities. During the pandemic, we saw the power of social media with its ability to start a revolution of change through campaigns like Black Lives Matter. Governments used social media platforms to educate their citizens on the covid 19 pandemic. Through such initiatives championed by individuals, we have seen the benefits of what social media can do for society.

**Social benefits**
Using social media can have a lot of positive social effects. Due to the ability to interact with others who have similar interests and experiences, social media may increase social ties and lessen loneliness, according to some research. Social media may be a helpful tool for sustaining and creating online groups as well as for expressing oneself. Johnson, T. J., & Kaye, B. K. (2014) states that Social media can have a variety of benefits on social relationships and well-being. For instance, it has been discovered that using social media is linked to higher levels of social capital (i.e., the resources that people may access through their interactions with others) and lower levels of loneliness. Evans, N. J., Schwartz, H. A., & Bos, J. E. (2017) In this study, a sample of college students' Facebook usage was compared to their overall well-being. According to the authors, Facebook use was connected with better mental health when it meant revealing more personal information and communicating with people close to you.

**Risk awareness**
The amount of research on the privacy risks connected to social media use is expanding. According to several important results from recent studies, numerous social media users are ignorant of the privacy dangers they encounter when using these platforms. Only 25% of social media users completely comprehend the privacy regula-tions of the platforms they use, according to a survey published in the Journal of the Association for Information Science and Technology Sheldon, (2020). Users of social media frequently share personal data in exchange for perceived advantages like access to exclusive material or tailored recommendations.

According to a study in the journal Computers in Human Behavior Acquisti & Grossklags (2005), people are more inclined to divulge personal information on social media when they anticipate good results.

Social media firms frequently capture and use people's personal information for profit by using data mining and targeted advertising strategies. A study indicated that social media businesses frequently use user data for targeted advertising, potentially violating users' privacy. The study was published in the journal Information, Communication & Society Mossberger, Tolbert, & McNeal (2013). Social media corporations' usage of personal information is neither transparent nor under our control. According to a study in the journal New Media & Society, social media users frequently have very little power over and little knowledge of how their personal data is utilized (Lenhart, 2015). We investigated previous research and came to the conclusion that these are the primary risks users are aware of.

**Personal risks**
People were one of the main risks that kept showing up in our data collection and from existing research. While people can be a great benefit they can also be a great risk. Users mentioned that not being aware of and enforcing their privacy settings could increase the likelihood of their posts being scrutinized by their followers. According to research, social media users frequently divulge personal information such as their home address, phone number, and location data, putting them in danger of identity theft and other privacy infractions Acquisti & Gross, (2006). If this is not controlled, Studies have found a link between using social media and bad mental health problems like depression, anxiety, and low self-esteem Baumeister & Leary, (1995); Kross et al., (2013).

**Security risks**
There is no disputing that social media has altered the way we communicate with one another. With billions of individuals globally, social media platforms have become a platform for exchanging information, ideas, and personal experiences. When it comes to privacy, though, social media use can be hazardous. In recent times we have seen how social media has led to the mass destruction of people's lives etc. We have seen young people destroy their reputations by posting the wrong content and people get stalked by the mere fact that they posted too much and gave too much and brought unnecessary attention to themselves. Krasnova, Wenninger, Widjaja, and Buxmann (2010) found that while users of social media platforms were aware of privacy hazards, they frequently did not appreciate the entire scope of these risks. According to Brandtzaeg, P. B., & Heim, J. (2012), social media users have limited privacy awareness, and most users trust social networking sites to secure their personal information.

**Social risks**

In terms of social risk. The main focus here will be on the user's inability to regulate how their content is used once it has been shared online. We have seen cases of impersonation and defamation from the content being reshared on social media by third parties. A person's post can be shared without their knowledge whether they have a private or public account. Screenshots can be taken and viewers can misunderstand the posts in themselves. During our data collection, one of the social risks we accessed was users' awareness of knowing their posts can be misinterpreted or misunderstood by others and this was a point most strongly agreed with.

## 2.3   Self-disclosure

Sharing habits in this literature review refer to the practice of users sharing their information on social media. In Levy, Gudes, & Gal-Oz (2018), they argue that although social media platforms provide room for their users to control their privacy, most users are unaware that the private information they share might leak to the users they do not wish to share it. Recently, sharing information on social media has been likened to being creative. It has even led to jobs like content creation, social media influencing, and social media marketing. All these roles require that content is shared on a constant basis and has even been identified as a way of increasing followers and likes on a social media account(Twitter and Instagram) Chua & Chang, (2016) This has not only transitioned to businesses but has also caused individuals to feel obliged to share their content and live their lives on social media as a way of increasing their followers and likes. From Chua & Chang (2016), it was identified that young people share selfies and post their daily activities online because they do not want to feel left out from their peers. Their sole aim is peer recognition.

Another important point is the fact that users did not mind posting false information as long as it made them feel recognized. Also, some users used fake accounts or provided false information in the creation of their accounts and hence do not see anything wrong with sharing the information on the platform. It was discovered by Krasnova, Spiekermann, Koroleva, & Hildebrand (2010) that users are motivated to share information due to the ease of social media platforms. This research focuses on understanding the factors that influence user privacy-sharing behaviors on social media.

   Self-disclosure refers to intentionally and voluntarily revealing information about oneself to others (Derlega and Grzelak, 1979). Self-disclosure is a common form of communicative behavior to share personal information, express oneself, or build social ties. The disclosure level varies with the degree of intimacy, precision, and awareness Altman and Taylor, 1973, Wheeless and Grotz, (1976). As social media is designed to encourage users to voluntarily generate content Boyd and Ellison, (2007), users express their opinions, beliefs, and moods and share personal details within the platform Krasnova et al., (2012). These activities' involvement results from the users' self-disclosure on social media.

   Social media users have various motivations to self-disclose, including building and maintaining relationships, increasing popularity, self-expression, self-clarification, and social entertainment Bazarova and Choi, 2014, Kashian et al.,( 2017), Ledbetter et al., 2010, Utz,(2015). These motivations can be interpreted as the benefits of disclosure on social media, which help users form positive attachments (e.g., feelings of enjoyment) to these platforms VanMeter et al., (2015). Empirical studies have revealed various forms of anticipated or perceived benefits, which can increase the willingness to disclose, including trust beliefs and enjoyment Krasnova et al., (2012), convenience Dinev et al., (2013), financial compensation Xu et al., (2009), or general benefits of social media use Dienlin and Metzger,( 2016).

The term "privacy paradox" refers to the discrepancy between people's privacy concerns and their behavioral patterns, like sharing private information online. This discrepancy has been seen in a variety of settings, such as social media, internet shopping, and managing health data. The social compensation hypothesis, which states that individuals may publicly reveal more private information because of the perceived advantages of community connection and support, has been applied to explain the privacy paradox. This theory is backed up by research that demonstrates that individuals who are lonely or have low self-esteem are more inclined to engage in self-disclosure on social media Dienlin & Trepte, (2016). This research considered three common benefits and risks users identified and is measuring users' self-disclosure based on the knowledge of their awareness in reference to their tendency to self-disclose.

By the end of this research, these are the two main questions we hope to answer:

1. How do younger users feel about the benefits and risks of disclosing themselves on social media?

2. How is the younger users' self-disclosure influenced by their awareness of privacy risks and benefits?

We will be testing four main hypotheses with the data we collect.

H1: There is a main effect of benefit awareness on social media user self-disclosure.

H2. There is a main effect of risk awareness on social media user self-disclosure.

H3: There is a main effect of different social media platforms on social media users' self-disclosure.

H4: There are interaction effects on risk awareness, benefit awareness, and social media platforms on users' self-disclosure

## 3  Methods

### 3.1  Participants and data collection

Two Hundred and sixty students from the University of Cincinnati ranging from ages 18 to 30 voluntarily participated in this survey. A group of students were awarded extra credit for participating in this survey while others were encouraged to participate. This decision was solely based on the discretion of their professors to encourage participation. Informed consent was obtained from all participants.

A survey was created using Microsoft forms to collect participant data and establish consent. A five-point Likert scale was used to measure participants' level of agreement with the survey questions. All surveys were conducted on a computer with the results extracted and analyzed with SPSS.

### 3.2   Measurements

The survey used for data collection consisted of univariate instruments. The main scale for the measurement of people's risk and benefit awareness level and their self-disclosure tendency was the 5-point Likert scale. There were seven statements to express privacy benefits, six for privacy risks, and ten for self-disclosure. To check the reliability of the statements used for the measurement in this study we embarked on a pilot study which had a couple more statements.

During the analysis of our pilot study, we removed items with low correlation(<0.5) and did a reliability analysis to remove items that contribute negatively to Cronbach's alpha value. For statements that had less than 0.5 correlation, we removed them from the survey.

### 3.3   Procedure

Participants were given a survey link or QR code via their canvas applications. Some professors gave their students extra credit for participating in the survey. This decision was solely based on the professor's discretion. The professors encouraged those not given extra credit to participate according to their capacity. The survey was open for a maximum of sixteen days in total. The description section of the survey briefly summarized the purpose of the research and participants' expectations. The link to the Informational consent form was embedded in the description section.

The main form had four distinct sections. The first section of the form was to reiterate the eligibility (age, willingness to voluntarily participate, and active participants of the social media platforms in the scope of the research) of desired participants. The second section was structured to understand participants' level of engagement on their preferred social media platform. Participants were asked to select their most used platforms and how often they interact and engage on a daily basis. The third section was to analyze their awareness level of the benefits and risks of sharing information on social media and their self-disclosure intention.

They were presented with statements on personal, social, and societal risks and benefits and possible reasons for self-disclosure and had to make a decision based on a five-point Likert scale.

After each subsection, they were asked open-ended questions to give their feedback on the benefits and risks and why they disclose information on social media.

The fourth and final section was on demographics. Here they were asked to select their age range and their level of education.

## 4 Results and Analysis

Of the 260 students we had, 115 participants which form 44% of our population is between the ages of 18 and 20 years old, 64 participants(25%) were between the ages of 21 and 22 years, 42 participants(16%) fall between the ages of 23 and 25 and 38 participants (15%) were 26 and above years old.

Regarding their academic level, we had a good span of participants across all levels from sophomore to graduate school. From the three main applications in the scope of this research, we had 123 participants(47%) choose Instagram as their most used application followed by Snapchat which had a total of 59 people(23%) report to be their most used application. 55 participants reported TikTok as their most used application, accounting for 21% of the total participants and 23 (9%) participants reported using other social accounts like Twitter, Facebook, Whatsapp, and discord.

The two research questions and three hypotheses were tested with factor analysis and a three-way analysis of variance(ANOVA). Independent variables were privacy benefit awareness, privacy risk awareness, and social media platforms while self-disclosure was our dependent variable.

### 4.1 Exploratory factor analysis

We measured participants' level of privacy awareness of the benefits and risks of using social media by providing statements and giving them options to respond to with a 5-point Likert scale. These statements covered the three types of risks and benefits we believe people perceive from their use of social media, which has been discussed in the literature review. What we saw in our reports was that there are two main components that influence younger users' benefits and risks awareness. We interet this as two main types of benefit and risk awareness. The KMO value for sampling adequacy for benefit awareness is .830 which means that the variables share a lot of information with each other and that there is a strong partial correlation. As a result, factor analysis is feasible.

**Benefit Awareness**
From our survey, we saw that there are two main components that users perceive as benefits awareness. We saw that staying in touch with friends, sharing restaurants and spots as recommendations to friends, and sharing accomplishments were highly correlated(>0.5) and form one major component. They fall between personal benefits and social benefits and we can conclude that socio-personal benefits are one of the main types of benefit awareness. The lower three were also correlated, assuming they have something to do in common and we will name them societal benefits. The first four statements will be summed up as socio-personal benefits and represent component one in our pattern matrix table. Component two represents societal benefits which also show correlation but the figure is not significant enough(<0.5) hence we conclude they are the second type of benefit awareness people gain from social media.

From our screen plot, we see that the two factors have an eigenvalue greater than 1, confirming that two main components influence benefit awareness.

**Risk awareness**

Likewise, for risk awareness, we saw that two main components represent the types of risk awareness users perceive. We saw a high correlation(>0.6) between the responses for those who were more aware of the likelihood of getting impersonated, having their content shared without their knowledge, and having their post misunderstood by others. The KMO value for our risk awareness statements was .730, indicating adequate sampling. We noticed some of the risks also coincided and could be both personal and security simultaneously, so from our pattern matrix component one represents personal security risk and component two is social risks. We conclude that security and personal risks are the two main types of risk users are aware of for the highly correlated figures. Our screen plot confirms that these two components have an eigenvalue (>1), further solidifying our conclusion that there are two main types of risk awareness.

## 4.2 Analysis of Variance

From the Test of Between subject effects table below we see that there is a main effect of benefit awareness on the self-disclosure of users on social media. Therefore we conclude that the more users are aware of the benefits of social media they will disclose information on social media(F= 36.291; df 1; $p < .001$) hence hypothesis one is accepted.

Our data also shows that there is a main effect of risk awareness recorded on self-disclosure (F= 7.001; df 1; $p < 001$). We conclude that users who are more aware of the risks of social media share less information on social media and those who are less aware of the benefits tend to share more on social media hence hypothesis two is accepted.

We see that the social media platform does not have a main effect on users' self-disclosure(df =3, F=.635, n/s ). This means that people do not self-disclose information differently based on their social media platform. This could be because the platforms being investigated have a similar mode of operation. They are heavy on pictures and videos instead of text, which could be the main reason for this finding, and based on this we reject the third hypothesis. And will recommend further investigation for future study.

We see that there is no main effect between the type of social media platform used, users' benefit awareness, and their risk awareness(df = 3, F= .715, n/s) which means that people do not share or share information based on the platform they use, the benefits they receive or the risk they are aware of. We also recorded that there is no main effect on self-disclosure based on the level of benefit and risk awareness users ha(df =1, F =0.021,n/s) which means that a person's decision to self-disclose is not based on only the benefits and the risks they are aware of. Our data also shows that there is no main effect on benefit awareness and social media platforms with regard to self-disclosure (df=3, F= .11, n/s) which means that people do not disclose information based on the social media app alone. The interactive effect of risk awareness and social media platforms was (df =3, F =2.658,p <0.05) which could mean that the risks on various social media platforms vary and based on the platform users measure their risk awareness differently before self-disclosing information. For this research further investigation was not done to probe further into which specific platforms this use case falls into but is something to look into in future studies.

## 5   Discussion and Conclusion

While we saw that there will be no significant impact on self-disclosure concerning benefit awareness and risk awareness, there will be a significant impact on self-disclosure concerning risk awareness and the social media platform(df 3, F=.715, sig > 0.5 ). This suggests that the risks on social media platforms vary to some extent. From the data collected, High benefit awareness and low-benefit awareness have a significant impact on self-disclosure. People who know more about the benefits they get from social media tend to share more, while those who know less about the benefits tend to share less on social media. Likewise, people who know more about the risks of sharing on social media tend to share less on social media while people who know less about the risks of sharing on social media share more on social media platforms. For the significance level of the third factor, a social media platform, we recorded a high significance figure, which allowed us to conclude that irrespective of the social media platform users sharing habits do not change much. Users who are more aware of the benefits they get from social media tend to share more and those who are less aware of the benefits tend to share less on social media. Out of the strongly disagree (1) to strongly Agree (5) the ability to stay in touch with friends recorded a mean response of 4.33. The responses for the first four questions in the survey which had to do with staying in touch with friends, sharing accomplishments for visibility, and sharing restaurants and vacation places with friends reported a high correlation of numbers close to 1 meaning these four statements which fall under personal and societal benefits strongly influence users to engage on their favorite social media applications.

On the contrary users, most of our participants reported that their engagement does not increase based on how their content is engaged or not, and sharing vacation pictures for others' experiences which falls under social benefit was not a strong benefit of engaging on social media. This answers that people engage on social media mainly to stay in touch with friends and for the purpose of the community.

On the flip side aside from people not wanting to be impersonated, the next high risk we recorded was people not being comfortable with others being able to post their content without permission. This statement recorded a mean response of 4.3. These mean recordings are from the 5-point Likert scale where 1 (Strongly agree ) and 5 (Strongly disagree)In this section of the survey, we had six statements that covered personal, security, and social risk. Amongst these statements the users getting misunderstood for their posts, posts being shared without consent, and fear of being impersonated or stalked had a mean value of 0.7 - 0.85. This statistic shows that peo-ple are concerned about their personal safety and how they are perceived by others on social media.

Benefit awareness and social media platforms also showed people's sharing behavior does not change based on the platform they use. However, we reported a low significant figure for risk awareness and user sharing on social media platforms which suggests that the risk on the platforms varies and that users share less or more based on the platform they are using. For future studies, we will recommend further investigation into the social media platforms and their individual risks on self-disclosure.